\documentclass{emulateapj}

\shorttitle{GRB afterglow plateaus and Gravitational Waves}
\shortauthors{A. Corsi and P. M\'esz\'aros }

\begin{document}

\title{GRB afterglow plateaus and Gravitational Waves: \\multi-messenger signature of a millisecond magnetar?}

\author{Alessandra Corsi\altaffilmark{1,2,3} and Peter M\'esz\'aros\altaffilmark{2} }

\email{alessandra.corsi@roma1.infn.it}

\email{nnp@astro.psu.edu}

\altaffiltext{1}{Universit\`a di Roma ``Sapienza'' and INFN-Roma, Piazzale Aldo Moro 2, 00185 - Roma (Italy)}
\altaffiltext{2}{Department of Astronomy \& Astrophysics, Department of Physics, and 
Institute for Gravitation and the Cosmos, The Pennsylvania State University, 
University Park, Pennsylvania 16802 (USA)}
\altaffiltext{3}{IASF-Roma/INAF, Via Fosso del Cavaliere 100, 00133 - Roma (Italy)}

\begin{abstract}
The existence of a shallow decay phase in the early X-ray afterglows of gamma-ray
bursts is a common feature. Here we investigate the possibility that this 
is connected to the formation of a highly magnetized millisecond pulsar, pumping 
energy into the fireball on timescales longer than the prompt emission.
In this scenario the nascent neutron star could undergo a secular bar-mode 
instability, leading to gravitational wave losses which would affect the 
neutron star spin-down. In this case, nearby gamma-ray bursts with isotropic
energies of the order of $10^{50}$~ergs would produce a detectable gravitational
wave signal emitted in association with an observed X-ray light-curve plateau,
over relatively long timescales of minutes to about an hour. The peak amplitude
of the gravitational wave signal would be delayed with respect to the gamma-ray 
burst trigger, offering gravitational wave interferometers such as the advanced 
 LIGO and Virgo the challenging possibility of catching its signature on the fly.
\end{abstract}


\section{Introduction}
\begin{figure}
\begin{center}
\includegraphics[width=8cm]{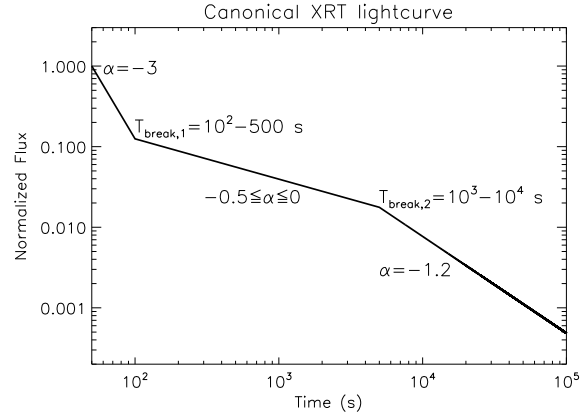}
\caption{Cartoon representation of the typical light curve behavior observed by \textit{Swift} XRT.
The ``standard'' power-law decay with index $\alpha=-1.2$ is preceded by a flat phase,
lasting $10^2-10^4$~s, during which the decay index is $\alpha=-0.5$ or flatter \citep{Zhang06}.
\label{canonical}}
\end{center}
\end{figure}

Thanks to \textit{Swift} observations \citep[e.g. ][]{Nousek06,Zhang06}, it has now 
become evident that the ``normal'' power-law behavior of long GRB X-ray light 
curves $F(T)\propto T^{\alpha}$ with $\alpha \sim -1.2$ (where $F(T)$ is the 
observed flux and $T$ is the observer's time), is often preceded at early times by 
an initial steep decay ($\alpha\sim-3$), followed by a shallower-than-normal decay 
($\alpha \gtrsim -0.5$, see Fig. \ref{canonical}). The steep-to-shallow and 
shallow-to-normal decay transitions are separated by two corresponding break times, 
$100~{\rm s}\la T_{break,1}\la 500$~s and $ 10^{3}$~s~$\la T_{break,2}\la 10^{4}$~s. 
During the shallow-to-normal transition the spectral index does not change 
and the decay slope after the break ($\alpha \sim -1.2$) is generally
consistent with the standard afterglow model \citep[e.g. ][]{Mes97,Sari98},
while the decay slope before the break is usually much shallower. The lack of
spectral changes suggests that the shallow phase may be attributed to a continuous
energy injection by a long-lived central engine, with progressively
reduced activity \citep[for a review see ][ and references therein]{Zhang06}.
Recently, \citet{PanaitescuV08} have pointed out that the effects of a late-time
energy injection may also be evident in some optical afterglows, around $30 -
10^{4}$~s after the trigger. Although it is still not clear if a typical
``steep-flat-steep'' behavior does exist also in short GRB X-ray afterglows, the
case of GRB~051221a does fit this scheme remarkably, with a plateau observed right
in the middle of the afterglow decay \citep{Soderberg06}.

Newborn magnetars are among the progenitors proposed to account for shallow decays 
or plateaus observed in GRB light curves \citep[][]{Dai98,Zhang01,Fan06,Yu07}.
Independent support for this scenario comes from the observation of SN2006aj, 
associated with the nearby sub-energetic GRB~060218, suggesting that the 
supernova-GRB connection may extend to a much broader range of stellar masses than
previously thought, possibly involving two different mechanisms: a ``collapsar''
for the more massive stars collapsing to a black hole (BH), and a newborn neutron star
(NS) for the less massive ones \citep{Mazzali06}.

Previous studies aimed at accounting for the afterglow plateaus by invoking a
magnetar-like progenitor have assumed that the magnetar's slow-down is dominated
by magnetic dipole losses, neglecting the contribution from the emission of 
gravitational waves \citep[GWs, see][]{Dai98,Fan06,Yu07}, or treating them separately
as a limiting case for a NS with sufficiently high, constant eccentricity
\citep{Zhang01}. Such studies have shown how magnetars dipole losses may indeed
explain the flattening observed in GRB afterglows. In the simplest version of the 
magnetar scenario, the end of the shallow decay is accompanied by an achromatic
break, while several cases of chromatic breaks have \textit{also} been observed 
\citep[e.g.][]{Panaitescu09}.  Additional mechanisms such as variable micro-physical 
parameters in the fireball shock front \citep[e.g.][]{Panaitescu06} or a structured jet model \citep[e.g. ][]{Racusin08} 
can be invoked to explain such chromatic breaks. Anyhow, a larger sample of 
simultaneous optical-to-X-ray observations is needed to firmly asses the achromatic 
or chromatic behavior of breaks associated to the end of the shallow-decay phase.

In this paper, we investigate in more detail the effects of GW losses on the 
magnetar's spin-down, and explore quantitatively the signatures which
could test whether this is indeed the mechanism at work in the shallow
X-ray light curves.
Although the precise evolution of a newborn magnetar from birth up to timescales 
of $\sim 10^3-10^4$~s is difficult to predict or to follow with numerical simulations,
here we point out that among the possible evolutionary paths which one may reasonably 
consider, one plausible and particularly interesting possibility to explore is that 
of a newborn NS left over after a GRB explosion, which undergoes a secular bar-mode 
instability.  In this scenario, simple estimates accounting for the most relevant 
energy loss processes can provide useful insights into the viability of having 
efficient GW emission associated with a GRB X-ray afterglow plateau. Although 
these estimates are clearly approximate, since possible complications like 
viscosity effects or magnetic field driven instabilities are neglected, they 
nonetheless allow us to make a first statement on the relevance of the considered process. 
Moreover, while other scenarios are also possible, the interesting aspect of this particular one is
that, on the one hand, GW observations would be facilitated by the presence of an 
electromagnetic signature to pinpoint the GW signal search, while on the other 
hand the detection of bar-mode like GWs in coincidence with a GRB X-ray plateau 
would be a smoking-gun signature of a magnetar pumping energy into the fireball,
thus identifying the much-debated plateau mechanism.
Given that several alternative scenarios have been invoked to explain the afterglow
flattening, which are \textit{not} expected to be associated with GW signals
\citep[see e.g. ][]{Panaitescu08}, this would represent a significant step forward
in our understanding of GRB physics. Moreover, identifying the presence of a magnetar 
would confirm that not all GRB explosions necessarily lead to the prompt formation 
of a BH.

A point of interest for current analyses that GW detectors are carrying out
\citep[see e.g. ][]{VirgoGRB,LigoGRB,LigoGRBshort} is that the scenario described here involves
a new class of GW signals, which should be searched for in coincidence with GRBs.
These would have a longer duration ($10^3-10^4$ s) and a different
frequency evolution than the type of GW signals currently considered to be possibly associated 
with  GRBs. 
Moreover, being delayed by minutes to $\lesssim$ 1 hour with respect to the prompt  $\gamma$-ray 
trigger, the GW signal associated with a GRB plateau would offer the challenging 
possibility of an on-line detection. 
In light of the fact that the Virgo\footnote{www.virgo.infn.it} and
LIGO\footnote{www.ligo.clatech.edu} interferometers are now progressing toward
their enhanced/advanced configurations, and getting prepared for performing
on-line data analyses, this prospect appears very appealing. It is worth noting
that despite a GW signal in coincidence with a GRB plateau could also be searched
off-line by LIGO or Virgo, an on-line detection would be highly preferable,
since it could serve as a trigger for ground-based optical follow-ups,
even if a GRB trigger alert is absent for any reason.

The paper is organized as follows. In Sec. \ref{sec1} we briefly describe how
GRB afterglow plateaus are modeled in the context of the magnetar model. In Sec.
\ref{sec2} we review the main processes that can lead to GW emission associated
with NS formation. The aim of this section is to show that, among the different 
mechanisms that can come into play, the high efficiency of the secular bar-mode 
instability is conducive to producing GW signals which are detectable also from 
relatively nearby extra-galactic sources. Moreover, it develops on timescales 
compatible with the observed durations of GRB plateaus. Sec. \ref{sec3} describes 
the general idea and particular aspects of the scenario being explored here, 
and how it can explain GRB afterglow plateaus with the presence of a 
magnetar whose spin-down includes both magnetic dipole and bar-mode GW losses. 
In Sec. \ref{sec4} we present the results of our calculations,
and in Sec. \ref{sec5} we discuss these results, summarizing our conclusions
in Sec. \ref{sec6}.

\section{GRB plateaus in the magnetar scenario}
\label{sec1}
Although a wide range of GRB progenitors end in the formation of a BH-debris torus system,
it has been proposed that some progenitors may lead to a highly magnetized rapidly
rotating pulsar \citep[e.g. ][]{Usov92,Duncan92,Thompson94,Yi98,Blackman98,Dai98,Kluzniak98,Nakamura98,Spruit99,Wheeler00,Ruderman00,Levan06,Mereghetti08,Bucciantini09}, with such possibility being realized not only in the case of long GRBs associated to collapsars,
but eventually also in scenarios relevant for short GRBs, such as NS binary mergers
\citep[][ and references therein]{Dai98}.

Fast rotating highly magnetized pulsars, are among the class of progenitors that
may be associated with significant energy input in the fireball for timescales
longer than the $\gamma$-ray emission, thus being relevant for
explaining GRB afterglow plateaus. A detailed analysis of the observable effects
linked to the presence of a pulsar pumping energy into the fireball was performed by
\citet[][]{Zhang01}, the results of which we briefly recall in what follows.

Consider the general scenario where the GRB is powered by a central engine that
emits both an initial impulsive energy input, $E_{imp}$, as well as a continuous
luminosity, the latter varying as a power-law with time, i.e. $L = L_0 \left(\frac{T}{T_0}\right)^{q}$
where $T$ is the observer's time. This could be the case if the central engine is a pulsar and
the initial impulsive GRB fireball is due to $\nu$-$\overline{\nu}$ annihilation or
magnetohydrodynamical processes \citep[see e.g. ][]{MacFadyen99,Popham99,DiMatteo02,Lee05,Oechslin06,Zhang09}.
In such a case, a self-similar blast wave is expected
to form at late times. In a GRB, the timescale $T_0$ at which the
self-similar solution applies is roughly equal to the time for the external
shock to start decelerating while collecting material from the interstellar medium
\citep[e.g. ][]{Sari1999}. At different times, the total energy into the
fireball may be dominated by either the initial impulsive term, or by the
continuous injection one, whose contribution will scale as
$E_{inj}=\frac{L_0 T_0}{q+1} \left(\frac{T}{T_0}\right)^{q+1}$. The continuous energy injection
term can dominate on the impulsive one for $T\gtrsim T_c$ (where $T_{c}\gtrsim T_0$
so as to assure that the self-similar solution has already developed when the
continuous injection law dominates), if $q>-1$ and $E_{inj}(T_c) \sim E_{imp}$.
In the particular case in which $L_0 T_0 \sim E_{imp}$ then $T_c \sim T_0$ and
the dynamics is dominated by the continuous injection as soon as the self-similar
evolution begins. Generally speaking, one can write $T_c=max\left\{T_0, T_0\left[(q+1)E_{imp}/(L_0T_0)\right]^{1/(1+q)}\right\}$ \citep[][]{Zhang01}.
Note that the continuous injection may, in addition, have another characteristic
timescale $T_{f}$ at which the continuous injection power-law index $q>-1$ switches
to a lower value $q<-1$. In such a case, it is only for $T_c < T_{f}$ that the
continuous injection has a noticeable effect on the afterglow light curve \citep{Zhang01}.

During the energy-injection dominated phase, the peak flux, peak frequency and cooling
frequency of the synchrotron photons produced by the forward shock \citep{Sari98} scale
with time as $F_m\propto T^{1+q}$, $\nu_m\propto T^{-(2-q)/2}$, $\nu_c\propto T^{-(q+2)/2}$,
respectively \citep{Zhang01}, that reduce to the standard scalings for $q=-1$ \citep{Sari98}. 
In the case of a nearly constant
energy supply, i.e. $q\sim 0$, one has $F_m\propto T$, $\nu_m\propto T^{-1}$, $\nu_c\propto T^{-1}$, respectively. These scalings allow one to compute the temporal indices of the afterglow
light curve expected during the injection phase. Supposing to be in slow cooling,
these are $F_{\nu}\propto F_m\nu_m^{(p-1)/2}\propto T^{\alpha_1}=T^{(3-p)/2}$ for $\nu_m<\nu<\nu_c$ and
$F_{\nu}\propto F_m\nu_m^{(p-1)/2}\nu_c^{1/2}\propto T^{\alpha_2}=T^{(2-p)/2}$ for $\nu>\nu_c$,
where we have indicated with $p$ the power-law index of the electron energy distribution in the shock front \citep{Sari98}.
For $2<p< 4$, one has $0.5>\alpha_1>-0.5$ at frequencies $\nu_m<\nu<\nu_c$ and $0>\alpha_2>-1$ at $\nu>\nu_c$,
to compare with $\alpha \gtrsim -0.5$ observed during GRB afterglow plateaus. In the
absence of energy injection, for the standard adiabatic fireball one would have
$-3/4>\alpha_1>-9/4$ for $\nu_m<\nu<\nu_c$, and $-1>\alpha_2>-5/2$ at $\nu>\nu_c$, for the same range of $p$ values.
Thus, the presence of a
pulsar pumping energy into the  fireball at a nearly constant rate, is expected
to cause a flattening in the typical decay of the afterglow light curve, with
$\alpha \gtrsim -0.5$, in agreement with \textit{Swift} observations (see Fig. \ref{canonical}).

\section{GWs by NS formation}
\label{sec2}
Gravitational collapse leading to the formation of a NS has long been considered
an observable source of GWs. During the core collapse, an initial signal is expected
to be emitted due to the changing axisymmetric quadrupole moment. A second part of the
GW signal is produced when gravitational collapse is halted by the stiffening of the
equation of state above nuclear densities and the core bounces, driving an outwards
moving shock, with the rapidly rotating proto-neutron star (PNS) oscillating in its
axisymmetric normal modes. In a rotating PNS, non-axisymmetric processes can also yield
to the emission of GWs with high efficiency. Such processes are convection inside
the PNS and in its surrounding hot envelope, anisotropic neutrino emission, dynamical
instabilities, and secular gravitational-radiation driven instabilities, that we
briefly recall in what follows. We refer the reader to e.g. \citet{Kokkotas08} for
a recent, more detailed review.

- \textit{Convection and neutrino emission - } 2D simulations of core collapse
\citep{Mueller2004} have shown that the GW signal from convection significantly
exceeds the core bounce signal for slowly rotating progenitors, being detectable
with advanced LIGO for galactic sources. In many simulations, the GW signature
of anisotropic neutrino emission has also been considered \citep{Epstein78,Burrows96,Mueller97}
and estimated to be detectable by advanced LIGO for galactic sources.

- \textit{Dynamical instabilities - } They arise from non-axisymmetric perturbations
and are of two different types: the classical bar-mode instability and the more recently
discovered low-$T/|W|$ bar-mode and one-armed spiral instabilities. In Newtonian stars,
the classical $m=2$ bar-mode instability is excited when the ratio $\beta=T/|W|$ of the
rotational kinetic energy $T$ to the gravitational binding energy $|W|$ is larger than
$\beta_{dyn}=0.27$ \citep{Ch69}. It can be excited in a hot PNS a few ms after core bounce, or alternatively
a few tenths of seconds later, when the PNS cools due to neutrino emission and contracts
further, with $\beta$ becoming $\gtrsim \beta_{dyn}$ ($\beta \propto 1/R$ during contraction).
The instability grows on a dynamical timescale
(the time that a sound wave needs to travel across the star) which is about one rotational
period, and may last from 1 to 100 rotations depending on the degree of differential rotation \citep[e.g.][]{Baiotti2007,Manca2007}.
 If the bar persists for $\sim$ 10-100 rotation periods, then even signals from
 distances considerably larger than the Virgo Cluster are estimated to be detectable.
An $m=1$ one-armed spiral instability has also been shown to become unstable in PNS, provided that the
differential rotation is sufficiently strong \citep[with matter on the axis rotating at least
ten times faster than matter on the equator,][]{Centrella2001,SBM2003}. In recent simulations
of rotating core collapse to which differential rotation was added \citep{Ott05}, the emitted
GW signal reached a maximum amplitude comparable to the core-bounce axisymmetric signal,
after $\sim 100$ ms and at a frequency of $\sim 800$ Hz.

- \textit{Secular instabilities -} At lower rotation rates, a star can become unstable to secular non-axisymmetric instabilities, driven by gravitational radiation or viscosity. Secular GW-driven instabilities are frame-dragging instabilities usually called Chandrasekhar-Friedman-Schutz \citep[CFS,][]{Chandra70,Friedman78} instabilities. Neglecting viscosity, the CFS-instability is generic in rotating stars for both polar and axial modes.
In the Newtonian limit, the $l=m=2$ $f$-mode, which has the shortest growth time of all polar
fluid modes \citep[$1~{\rm s} \lesssim\tau_{GW}\lesssim 7\times10^4$ s for $0.24\gtrsim\beta\gtrsim0.15$, see ][]{LaiShapiro95},
becomes unstable when $\beta\gtrsim0.14$. The $f$-mode instability, also referred to as the secular
bar-mode instability, is an excellent source of GWs. In the ellipsoidal approximation, \citet{LaiShapiro95}
have shown that the mode can grow to a large nonlinear amplitude, modifying the star
from an axisymmetric shape to a rotating ellipsoid, that becomes a strong emitter
of GWs until the star is slowed-down towards a stationary state. This stationary state is a
Dedekind ellipsoid, i.e. a non-axisymmetric ellipsoid with internal flows but with a
stationary (non-radiating) shape in the inertial frame. During the evolution, the non-axisymmetric pattern radiates
GWs sweeping through the advanced LIGO/Virgo sensitivity window (from 1 kHz down to about 100 Hz),
which could become detectable out to a distance of more than 100 Mpc. Two recent hydrodynamical
simulations \citep[][in the Newtonian limit and using a post-Newtonian radiation-reaction, respectively]{Shibata04,Ou04}
have essentially confirmed this picture.

Among axial modes, the $l=m=2$ $r$-mode is an important member
\citep[see e.g. ][]{Andersson98,Friedman98,Lindblom98,Owen98,Lindblom02,Andersson01,Andersson03,Bondarescu09}. If the compact object is a strange star, such instability is predicted to persist for a few hundred years (at a low amplitude) and, integrating data for a few weeks, could yield to an effective amplitude $h_{\rm eff}\sim 10^{-21}$ for galactic signals,
at frequencies $\sim 700-1000$ Hz \citep{Kokkotas08}.

- \textit{Other magnetic-field related effects} - Finally, mechanisms different from rotational instabilities can be invoked as GW sources in newborn magnetars. E.g., in several scenarios the star's shape may be dominated by the distortion caused by very high internal magnetic fields \citep[e.g. ][]{Palomba01,Cutler02a,Arons03,Stella05,Da07,Da08}. GW signals produced by these kind of processes are typically estimated to be detectable by the advanced interferometers up to the Virgo Cluster (i.e. distances of the order of $20$ Mpc).

\section{The NS spin-down}
\label{sec3}

On the longer afterglow timescales that are of interest for the present work, the energy
injection into the fireball by a magnetar eventually surviving after the GRB explosion,
is expected to be mainly through electromagnetic
dipolar emission \citep[][]{Zhang01}. For what concerns GW losses,
in this work we focus on the secular bar-mode instability, given its high efficiency in the
production of GWs, and being its characteristic timescale $\tau_{GW}$ compatible with the one of GRB plateaus (see Sects. \ref{sec2} and \ref{sec3}).

As discussed in the previous section, a collapsing core rotating sufficiently fast
is expected to become non-axisymmetric when $\beta$ is sufficiently large.
Since a new-born NS can be secularly unstable but dynamically stable only if the rotation
rate of the pre-collapse core lies in a narrow range, and since during the collapse $\beta$
increases proportionally to $R^{-1}$, \citet{LaiShapiro95} considered 
more likely that the core becomes dynamically
unstable ($\beta>\beta_{dyn}$) following the collapse, provided the initial $\beta_i$ is not too small.
On a short dynamical timescale, such NS will evolve toward a nearly axisymmetric equilibrium state,
with $\beta$ decreasing below $\beta_{dyn}$, but possibly remaining above $\beta_{sec}$
\citep[see][ and references therein]{LaiShapiro95}. Due to gravitational radiation, the
nearly axisymmetric core (secularly unstable Maclaurin spheroid) will evolve into a
non-axisymmetric configuration (Riemann-S ellipsoid), on a secular dissipation
timescale $\sim \tau_{GW}$. While an initial dynamical unstable phase would possibly
produce a GW burst during the GRB, the secular evolution takes place on longer timescales,
thus being relevant for the shallow phase  ($100$~s~$\la T \la 10^4$~s) observed in GRB
afterglows (see Fig. \ref{canonical}). For this reason, in what follows we focus on the \textit{secular} bar-mode instability. It is worth noting, however, that also the presence of a bar-like GW burst from a \textit{dynamical} bar-mode instability, would provide a hint for a magnetar being formed in the GRB explosion. BH formation, in fact, is not expected to lead to strong quadrupole moments \citep[except if it is argued for blobs forming in the infall, see e.g.][]{Kob03}, and in any case a dynamically unstable magnetar would presumably give rise to a more regular signal.

Fully general relativistic axisymmetric simulations of rotating stellar core collapse in three spatial dimension, performed for a wide variety of initial conditions (rotational velocity profile, equations of state, total mass), indicate that the threshold $\beta=0.27$ for the onset of the classical dynamical instability  is passed if the the progenitor of the collapse is: (i) highly differentially rotating; (ii) moderately rapidly rotating with $0.01 \lesssim \beta_{i} \lesssim 0.02$; (iii) massive enough \citep{Shibata2005}. 
More recent numerical collapse simulations of rotating stellar iron cores to PNS have also provided an extensive set of post-bounce rotational configurations, allowing studies of the prospects for the development of non-axisymmetric rotational instabilities. E.g. \citet{Dimmelmeier2008} found that the rotational barrier imposed by centrifugal forces prohibits the spin-up necessary for the classical dynamical bar-mode instability, but a large subset of post-bounce models exhibits a $\beta$ above the secular instability threshold. Based on their results, \citet{Dimmelmeier2008} consider it unlikely that a PNS in nature develops a high-$\beta$ dynamical instability at or early after core bounce. While many of the PNS could theoretically reach $\beta_{dyn}$ during their subsequent cooling to the final condensed NS (if angular momentum is conserved), it is however considered more likely that the secular instability driven by dissipation or gravitational radiation back-reaction will set in first \citep{Dimmelmeier2008}. Still, three-dimensional simulations are necessary to provide conclusive tests of these predictions.

Under the hypothesis that a secular bar-mode
instability does indeed set in, in this work we follow the NS quasi-static  evolution
under the effect of gravitational radiation according to the analytical formulation
given by \citet{LaiShapiro95}. Such evolution can in principle be studied using the
full dynamical equations of ellipsoidal figures \citep{Ch69}, including gravitational
radiation reaction. However, since $\tau_{GW}$ is generally much longer than the dynamical
time of the star, the evolution is quasi-static, i.e. the star evolves along an
equilibrium sequence of Riemann-S ellipsoids.  Differently from what done by \citet{LaiShapiro95},
here we add in the energy losses the contribution of magnetic dipole radiation, under the hypothesis
that those will not substantially modify the dynamics, but will
act speeding up the overall evolution of the bar along the same sequence of Riemann-S ellipsoids
that the NS would have followed in the absence of radiative losses. As we are going to
show in the following section, dipole losses are nearly constant during the bar evolution
so that, according to what discussed in Sec. \ref{sec1} for the $q=0$ case, they can
act as a source of continuous energy supply into the fireball, explaining the observed
slope of GRB afterglow plateaus.

It is worth noting that in a real situation where magnetic field instabilities and viscosity effects
are also present, the relevant timescales may be altered.
A secularly evolving bar can last up to a timescale of the order of $10^3$ s, as far as viscosity or
magnetic field induced instabilities do not substantially modify the dynamics.
Viscosity may play a role on the secular evolution when the PNS has cooled to
below $\sim 1$ MeV \citep[e.g. ][]{LaiShapiro95,Lai01}. The time required for
the pulsar to cool below such temperature was estimated as a few hundreds of
seconds by \citet{Lai01}. In the case of a GRB explosion, a less rapid cooling
is expected due to continuous in-fall and jet emission \citep[a heating source
which was absent in e.g. ][]{Lai01}, so that we may assume that the bar survives
at least until the end of the electromagnetic plateau (i.e. $T\sim10^3$ s).
Magnetic effects are notoriously difficult to predict \citep[see e.g.][]{Shibata04},
and in general require making heuristic assumptions. 
In the context of the secular r-mode instability, \citet{Rezzolla2001} have shown that the growth of an initial magnetic field associated with the secular kinematic effects emerging during the evolution of the instability, possibly damps the growth of the instability itself. Despite the different context (r-modes), these results do suggest that magnetically driven instabilities may complicate the scenario.
In what follows, we explore the
quantitative consequences of making the plausible assumption that magnetic instabilities
are less efficient at spinning-down the bar than GW emission and magnetic dipole losses.
Moreover, apart from magnetic braking (spin-down due to dipole losses) which we do consider here, the presence of a magnetic field can influence the secular evolution in other ways. A magnetic field anchored on the star's surface is in fact perturbed by the instability itself, and this can lead to electromagnetic losses which can enhance the CFS mechanism. E.g. in the context of r-modes, \citet{Ho2000} have considered the electromagnetic radiation associated with the shaking of magnetic field lines by the r-mode oscillations. This effect has been estimated to be negligible for NS with magnetic field strengths below $10^{16}$ G. In our case, magnetic field lines anchored on the surface would be distorted by the bar-mode instability. In our treatment, we neglect this effect, and include only dipole losses associated with a magnetic field whose flux is conserved on a sphere of radius equal to the mean radius of the ellipsoid.

\section{Modeling of the NS evolution}
\label{sec4}
A general Riemann-S ellipsoid is characterized by an angular velocity $\Omega~\widehat{e}_3$
of the ellipsoidal figure (the pattern speed) about a principal axis $\widehat{e}_3$, and by
internal fluid motions which are assumed to have uniform vorticity $\zeta~\widehat{e}_3$ along the same axis
(in the frame co-rotating with the figure). Labeling with $a_1$ and $a_2$ the principal axes
of the ellipsoidal figure in the equatorial plane\footnote{In the ellipsoidal approximation, surfaces of constant density are assumed to be self-similar ellipsoids, so the geometry of the configuration is completely specified by the three principal axes $a_1$, $a_2$ and $a_3$, and the axis ratio $a_3/a_1$ and $a_2/a_1$ are the same for all interior isodensity surfaces \citep[][]{LRS93}.}, and with $x_1$ and
$x_2$ Cartesian coordinates in such a plane, it can be shown that the fluid velocity in the
inertial frame reads

\begin{equation}
	\vec{u}_0=\vec{u}+\Omega(\widehat{e}_3\times\vec{r})
	\label{eqvel}
\end{equation}

with

\begin{equation}
	\vec{u}=\frac{a_1}{a_2}\Lambda x_2 \widehat{e}_1 - \frac{a_2}{a_1}\Lambda x_1 \widehat{e}_2
\end{equation}

the velocity in the frame co-rotating with the figure \citep{Ch69}, where $\widehat{e}_1$ and $\widehat{e}_2$ are unit vectors along the Cartesian axes $x_1$ and $x_2$; $\vec{r}$ is the position vector; $\times$ indicates the vector product; and

\begin{equation}
	\Lambda=-\frac{a_1a_2}{a^2_1+a^2_1}\zeta
\end{equation}

is the angular frequency of the internal fluid motions, i.e. of the elliptical orbits that the particles span around the rotational axis in addition to the pattern motion.
The velocity $\vec{u}_0$ is contained in the plain perpendicular to the rotational
axis $\widehat{e}_3$, so indicating with $\vec{r}_{\bot}$ the projection of the position vector $\vec{r}$
in such plane we can write $\vec{r}=\vec{r}_{\bot}+x_3 \widehat{e}_3$, and $\vec{u}_0=d(\vec{r}_{\bot})/dt+(dx_3/dt)\widehat{e}_3=d(\vec{r}_{\bot})/dt$ (i.e. $dx_3/dt=0$ since the component of $\vec{u}_0$ along $\widehat{e}_3$ is null). Further, we can write $\vec{u}_0=(dr_{\bot}/dt)\widehat{r}_{\bot}+
r_{\bot}d(\widehat{r}_{\bot})/dt=(dr_{\bot}/dt)\widehat{r}_{\bot}+r_{\bot}(\vec{\Omega}_{0}\times \widehat{r}_{\bot})$,
where we have set 

\begin{equation}
	\vec{\Omega}_{0}=\frac{1}{r_{\bot}}(\widehat{r}_{\bot}\times \vec{u}_0).
\end{equation}

Note that $\vec{\Omega}_{0}$ is defined in such a way that $r_{\bot}\Omega_{0}$ 
gives the component of the particle velocity perpendicular to the polar 
radius $\vec{r}_{\bot}$, measured in the inertial frame. However, as underlined above, 
the motion of fluid particles on the surface can be viewed as the superposition of a 
circular motion with the pattern frequency $\Omega$, plus an elliptical motion on paths 
contained on the pattern ellipsoid (resulting in maintaining the pattern fixed). 
Since the internal fluid motions are ellipses rather than circles, there is an additional 
component of the velocity parallel to $\vec{r}_{\bot}$. Using Eq. (\ref{eqvel}), one has

\begin{equation}	\vec{\Omega}_{0}=\left[\Omega-\left(\frac{a^{2}_{2}x^{2}_{1}+a^{2}_{1}x^{2}_{2}}{r^{2}_{\bot}a_{1}a_{2}}\right)\Lambda\right]\widehat{e}_3
\end{equation}

In the frame co-rotating with the pattern, fluid particles on the star's surface move around the rotational axis, on ellipses contained in $x_3=const$ planes. Those ellipses are self-similar to the equatorial one and have equation:

\begin{equation}
	\frac{x^2_1}{a^2_1\left(1-\frac{x^2_3}{a^2_3}\right)}+\frac{x^2_2}{a^2_2\left(1-\frac{x^2_3}{a^2_3}\right)}=1
\end{equation}

For such fluid particles, $a^{2}_{2}x^{2}_{1}+a^{2}_{1}x^{2}_{2}=a^2_1a^2_2\left(1-\frac{x^2_3}{a^2_3}\right)$, and $\left<r_{\bot}\right>=\sqrt{a_1a_2\left(1-\frac{x^2_3}{a^2_3}\right)}$, so that in one cycle $\left<\frac{a^{2}_{2}x^{2}_{1}+a^{2}_{1}x^{2}_{2}}{r^{2}_{\bot}a_{1}a_{2}}\right>=1$. Thus, in the inertial frame, fluid particles on the star's surface are characterized by an angular frequency:

\begin{equation}
	\Omega_{eff}=\left<\Omega_{0}\right>=\Omega-\Lambda
\end{equation}

Since the gravitational radiation reaction acts like a potential force, the fluid circulation along the equator of the star,

\begin{equation}
	C=\int_{equator}\vec{u}_0\cdot d\vec{l}=\pi a_1 a_2 \zeta_0,
\end{equation}

where $d\vec{l}$ is taken along the star's equator and $\zeta_0$ is the vorticity in the
inertial frame, is conserved in the absence of viscosity \citep[][]{LRS93}. Therefore, the NS
will follow a sequence of Riemann-S ellipsoids with constant circulation. Treating the NS as
a polytrope of index $n$ \citep{Cha39}, total mass $M$, and indicating with $R_0$
the radius of the non-rotating, spherical equilibrium polytrope with same mass $M$, one has \citep[][]{LaiShapiro95}:

\begin{equation}
	\bar{\cal C}=\frac{{\cal C}}{\sqrt{G M^3 R_0}}=-\frac{M k_n C}{5\pi \sqrt{G M^3 R_0}}
\end{equation}

where $G$ is the gravitational constant; $k_n$
is a constant which depends on the index $n$ of the considered polytrope \citep[see e.g. ][]{LRS93}. Note that $\bar{\cal C}={\cal C}/{\sqrt{G M^3 R_0}}$ is an adimensional quantity, ${\cal C}=-(k_n M C)/(5\pi) $ has the dimensions of an angular momentum, and both are proportional to the conserved circulation $C$.
It can be shown that \citep[][]{LaiShapiro95}:

\begin{equation}
	{\cal C}=I \Lambda -\frac{2}{5} k_n M a_1 a_2 \Omega
\end{equation}

where $I=k_n M(a^2_1+a^2_2)/5$ is the NS moment of inertia with respect to the rotational axis. Along the secular equilibrium sequence, we write the NS spin-down law as \citep{ST}:
\begin{equation}
	\frac{dE}{dT}=-\frac{B^2_p R^6\Omega^4_{eff}}{6c^3}-\frac{32 G I^2 \epsilon^2 \Omega^6}{5 c^5}=L_{dip}+L_{GW}
	\label{eq1}
\end{equation}
where $E$ is the NS total energy, $L_{GW}=dE_{GW}/dT$ accounts for GW losses, while
$L_{dip}=dE_{dip}/dT$ for magnetic dipole ones. Here
$\epsilon=(a^2_1-a^2_2)/(a^2_1+a^2_2)$ is the ellipticity;
$B_p$ is the dipolar field strength at the poles; $\Omega$ is
the pattern angular frequency \textit{of the ellipsoidal figure}; $R$ is the
\textit{mean stellar radius}; $c$ is the light speed; $T$ is the time measured in an
inertial frame where the pulsar is at rest. $L_{dip}$ is computed
conserving the magnetic field flux over a sphere of radius equal to the mean stellar
radius (i.e. $B_{p}R^2=const=B_{p,0}R^2_0$ along the sequence, where $R$ is the geometrical mean of the
ellipsoid principal axes), and using the effective angular frequency $\Omega_{eff}$,
which includes both the pattern speed and the effects of the internal fluid motions. 
The use of $\Omega_{eff}=\left<\Omega_{0}\right>=\left<\frac{1}{r_{\bot}}|\widehat{r}_{\bot}\times 
\vec{u}_0|\right>$ accounts for the fact that in the frozen-in magnetic field approximation 
\citep[see e.g.][]{Goldreich69,Baym69,Thompson96,Morsink02,Thompson02} the magnetic field lines 
are in effect tied to the fluid particles on the stellar surface. Note that 
$\Omega_{eff}$ (and the corresponding dipole loss term) is measured in the inertial 
frame, which is where we compute $dE/dT$ as well.

Once $(\bar{\cal C},n,M,R_0,B_{p,0})$ are assigned, each configuration along a
constant-$\bar{\cal C}$ sequence is completely determined specifying the axis-ratio
$x=a_2/a_1$ in the ellipsoid equatorial plane. Thus, all relevant quantities can be
considered as functions of $x$ only, and Eq. (\ref{eq1}) can be written as:
\begin{equation}
	\frac{dx}{dT}=\frac{L_{dip}(x)+L_{GW}(x)}{dE/dx}
	\label{eq2}
\end{equation}

We solve the above equation numerically, with its right hand side evaluated along a
constant-$\bar{\cal C}$ Riemann-S sequence, and imposing an initial
condition sufficiently near to a uniformly rotating Maclaurin spheroid, ($x(t_i)=x_i
\rightarrow 1$) of the given circulation $\bar{\cal C }$.

\section{Results and Discussion}
\label{sec5}

\begin{figure}
	\begin{center}
		\includegraphics[width=8cm]{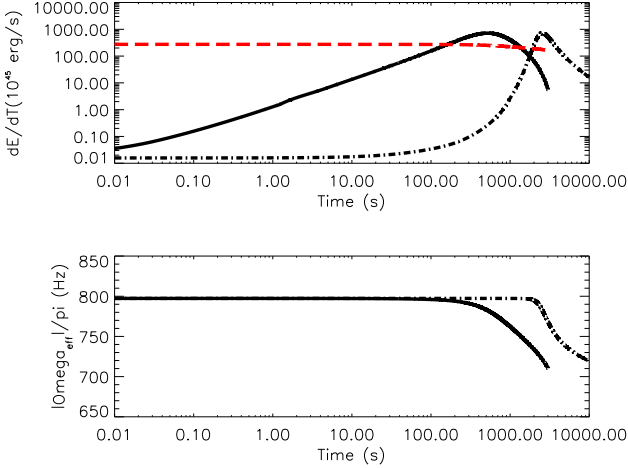}
		\caption{NS evolution along a Riemann-S sequence with parameters $(\bar{\cal C},n,M,R_0,B_{p,0})=(-0.41,1,1.4$
M$_{\odot}$, 20 km, $10^{14}$~G). Upper panel: Rate of energy loss in units of $10^{45}$ ergs/s, when both magnetic dipole losses (red-dashed line) and GW losses (black-solid line) are taken into account in the magnetar's spin-down law. For reference, we also plot the rate of energy loss in the case only GW emission is considered (black-dash-dotted line), as in \citet{LaiShapiro95}. Lower panel: absolute value of the surface fluid particles angular frequency divided by a factor of $\pi$ (i.e. $|\Omega_{eff}|/\pi$), when both magnetic dipole and GW losses are considered (black-solid line). For reference, we also plot the same quantity when only GW losses are taken into account in the magnetar's spin-down law (black-dash-dotted line), as in \citet{LaiShapiro95}. Note that the vertical axis in the lower panel is a linear scale: between $10^2$~s and $\sim 10^{3}$~s,  $\Omega_{eff}/\pi$ changes from $\sim 800$~Hz to $\sim 750$~Hz, i.e. less than $\sim 10\%$ of its initial value. Thus, between $10^2$~s and $10^3$~s the power-law approximation to dipole losses is $L_{dip}\propto T^{-0.11}$, so that $q\sim 0$ can be assumed for $T\lesssim 10^3$~s. (See the electronic version for colours).\label{fig1}}
		\end{center}
\end{figure}

In Fig. \ref{fig1} we compare the luminosity emitted in GWs, computed with (black-solid
line) or without (black-dash-dotted line) the addition of the dipole loss term (red-dashed line)
in Eq. (\ref{eq1}), for a typical choice of parameters, $(\bar{\cal C},n,M,R_0,B_{p,0})=(-0.41,1,1.4$
M$_{\odot}$, 20 km, $10^{14}$ G). Note that $\bar{\cal C}=-0.41$ corresponds to a value of $\beta=0.20$ for the initial Maclaurin configuration, i.e. in the middle of the $0.14<\beta<0.27$ range for the secular instability.
As evident from the lower panel of Fig. \ref{fig1}, as long as the circulation is conserved, $\Omega_{eff}$ remains nearly constant during all the evolution, and $|L_{dip}|\sim 3\times 10^{47} {\rm ergs/s}=L_0$ (upper panel, red-dashed line). As underlined in Sec. \ref{sec1}, energy pumped into the fireball at a constant rate is sufficient to explain the observed temporal behavior of afterglow plateaus (i.e. $\alpha \gtrsim -0.5$,
see Fig. \ref{canonical}). For what concerns the duration of the plateau, for a GRB with impulsive isotropic energy of the order of $E_{imp}\sim 10^{50}$ ergs, the effect of the energy injection in the light curve will become visible after a time $T_c\sim E_{imp}/L_0\sim (10^{50} {\rm ergs}) /(3\times 10^{47} {\rm ergs/s}) \sim 300$ s (see the red-dashed line in Fig. \ref{fig1} and Sec. \ref{sec2}), which is about in the middle of the observed range for $T_{break,1}\sim 100-500$~s (see Fig. \ref{canonical}). Supposing the energy injection ends or starts fading significantly when the star approaches the final Dedekind state (see the discussion at the end of this section), the GRB light curve will return to its standard behavior after $T_{break,2}\gtrsim 10^{3}$, to be compared with the observed range of $10^{3}-10^{4}$ s. Thus, for a GRB with such impulsive energy, the properties of the plateau associated to the NS secular evolution are in agreement with those typically observed\footnote{Note that in our discussion we are neglecting redshift effects, since we are interested in nearby GRBs at $z\lesssim 0.035$, i.e. having luminosity distances $d_L \lesssim 150$ Mpc.}.

\begin{figure}
	\begin{center}
		\includegraphics[width=8cm]{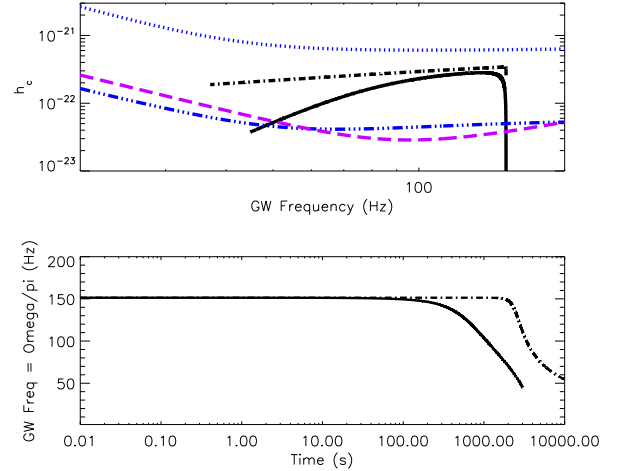}
		\caption{Upper panel: characteristic GW amplitude $h_c$ at $d=100$ Mpc, with dipole plus GW (black-solid line) and only GW \citep[black-dash-dotted line, see also ][]{LaiShapiro95} losses being considered. A typical fit to the sensitivity expected for advanced detectors \citep[purple-dashed line, see e.g.][]{Cutler94,Owen98}, Virgo nominal sensitivity (blue-dotted line), and the advanced Virgo sensitivity optimized for binary searches \citep[blue-dash-dot-dot-dotted line, ][]{Losurdo08}, are also shown. Lower panel: evolution of the GW signal frequency, with dipole plus GW (black-solid line) and only GW (black-dash-dotted line) losses being considered in the NS spin down. (See the electronic version for colours).}
		\label{fig2}
		\end{center}
\end{figure}

The waveform of the GW signal emitted in association with the afterglow plateau is computed as  \citep{LaiShapiro95}:
\begin{eqnarray}
\nonumber	h_+=-\frac{h(t)}{2}\cos\Phi(1+\cos^2\theta)~~~&h_{\times}=-h(t)\sin\Phi\cos\theta\\
	\label{GWsignal}
\end{eqnarray}
where $\theta$ is the angle between the line of sight and the rotation axis of the star, $\Phi=2\int^{t}_{t_0}\Omega t$ is twice the orbital phase, and
\begin{equation}
	h(t)=\sqrt{\left(\frac{2 c^3 d^2 \Omega^2}{5 G}\right)^{-1} |L_{GW}|}=\frac{4 G \Omega^2}{c^4 d}I\epsilon
	\label{eqampl}
\end{equation}
where $d$ is the distance to the source, $L_{GW}$ and $\Omega$ are shown in Fig. \ref{fig1} (upper panel, black-solid line) and Fig. \ref{fig2} (lower panel, black solid line), respectively. The resulting GW signal is quasi-periodic, with frequency $f=\Omega/\pi$.

To estimate the GW signal detectability, we proceed as follows.
For broad-band interferometers such as LIGO and VIRGO, the best signal-to-noise ratio
is obtained by applying a matched filtering technique to the data, when a waveform template is available. In such a case,

\begin{equation}
	\rho^2=4\int^{+\infty}_{0}\frac{(F^2_+ |\widetilde{h}_+ (f,\theta)|^2+F^2_{\times}|\widetilde{h}_{\times}(f,\theta)|^2)}{S_h(f)}~df
\end{equation}

where $\widetilde{h}$ is a Fourier transform; $S_h(f)$ is the power spectral density of the detector noise; $F_+$, $F_{\times}$ are the beam pattern functions \citep[$0<F^2_+, F^2_{\times}<1$ depending on the source position in the sky, see e.g. ][]{Thorne,Fla98}. For the signal in Eq. (\ref{GWsignal}), in the stationary phase approximation \citep{Thorne,Cutler94,Owen98,Owen02},

\begin{equation}
	\rho^2=\int^{+\infty}_{0}\frac{h^2(t)(dt/df)(F^2_+ (1+\cos^2\theta)^{2}/4 + F^2_{\times}\cos^2\theta)}{S_h(f)}~df.
\end{equation}

Since we expect to be observing the GRB on-axis, $\theta\simeq0$. In case of optimal orientation,

\begin{eqnarray}
	\nonumber \rho^2_{max}=\int^{+\infty}_{0}\frac{f^2 h^2(t)(dt/df)}{f S_h(f)} d(\ln f)=\\=\int^{+\infty}_{0} \left(\frac{h_c}{h_{rms}}\right)^{2} d(\ln f),
\end{eqnarray}

being $h_c=f h(t)\sqrt{dt/df}$ the characteristic amplitude, and $h_{rms}=\sqrt{f S_h(f)}$. In the upper panel of Fig. \ref{fig2}, we compare $h_c$ computed for a GRB at $d=100$~Mpc, with the $h_{rms}$ expected for the advanced detectors \citep{Losurdo08,Cutler94,Owen98}, for which $\rho_{max}\gtrsim 5$ at $d\lesssim 100$~Mpc, or $d\lesssim 150$~Mpc if we make the assumption that knowledge of the GRB trigger time reduces the detection threshold, of a factor which as a rule-of-thumb we take equal to $1.5$ \citep{Piran93,Cutler02}. Higher confidence in an eventual detection may require $\left\langle \rho\right\rangle_{sky}=\sqrt{2/5}~\rho_{max}\gtrsim 5$ in each of a three-detectors network with similar $h_{rms}$ \citep{Cutler94}. With the help of the GRB trigger time roughly compensating the factor of $\sqrt{2/5}$, this implies $d\lesssim 100$~Mpc.

BATSE results show that about $3\%$ of short GRBs are expected to be within 100~Mpc
\citep{Nakar2006}, which translates into $\sim 1-2$ short GRBs per year in the Swift
($\sim$ 10 short GRBs per year) plus GBM (GLAST - $\sim 1/4$ of $\sim$ 200 GRBs per year)
sample. As far as low-luminosity long GRBs, two of them (980425 and 060218) were already
observed at a $d\sim 40$ Mpc and $d\sim 130$ Mpc, and their local rate ($\gtrsim 200$
Gpc$^{-3}$ yr$^{-1}$) is expected to be
much higher than that of normal bursts \citep[1 Gpc$^{-3}$ yr$^{-1}$, e.g.][]{Virgili2009}.
INTEGRAL has detected a large proportion of
faint GRBs inferred to be local \citep{Integral}, a sample which may be increased by
future missions such as Janus \citep{Janus} and EXIST\footnote{http://exist.gsfc.nasa.gov/}. To compare with the case discussed here, the standard progenitor scenario for \textit{long} GRBs predicts $\rho\sim5$ at $27$~Mpc for the advanced Virgo/LIGO, while the chirp signal from \textit{short} GRBs is estimated to be detectable up to
several hundreds Mpc \citep[e.g.][]{Fla98,Kob03}. GWs eventually
detected after a chirp and
during an electromagnetic plateau of a short GRB, would add a significant piece
of information, probing whether a magnetar is formed in the coalescence, rather than a BH.

It is finally worth adding some few considerations more. First, in the scenario we are proposing here, some correlations do exist between the electromagnetic plateau and the GW signal, that could be explored in future analyses so as to test up to which level those may help the GW signal search. For example, a measurement of the initial frequency of the GW signal, for a given NS mass and radius, would allow one to derive an estimate $\beta_{GW}$ for the actual value of $\beta$ \citep[see also Fig. 5 in][]{LaiShapiro95}. In the ellipsoidal approximation, $\beta_{GW}$ would predict a specific evolution of the bar, as  e.g. the expected value of $\Omega_{eff}(\beta_{GW})=\Omega(\beta_{GW})-\Lambda(\beta_{GW})$ during the nearly constant phase. At the same time, the luminosity of the afterglow plateau, for a given NS mass, radius and magnetic field strength, would also allow one to estimate the value of $\Omega_{eff}$ during the constant phase, which could thus be checked for consistency with the value $\Omega_{eff}(\beta_{GW})$ inferred from the GW measurements. 

Next, some considerations are required on the fate of the bar after the final Dedekind state is reached. In the absence of dipole losses, the evolution of the NS
along the sequence would have maintained $\Omega$ nearly constant up to a time $T_{GW}$, of the order of few secular growth times, $\tau_{GW}\simeq 2\times 10^{-5}~{\rm s} \left[M/(1.4 M_{\odot})
\right]^{-3}\left[R_0/({\rm 10 km})\right]^{4}(\beta-\beta_{sec})^{-5}$ \citep{LaiShapiro95}.
Here $\beta$ is referred to the initial
Maclaurin configuration, and it's determined by the choice of $\bar{\cal C}$. In our case,  $\bar{\cal C}=-0.41$ and
$\beta=0.20$, so that $\tau_{GW}\simeq 335$~s. As evident from the black-dash-dotted line in the lower panel of Fig. \ref{fig2}, for such value of the circulation, when only GW losses are considered, one has $T_{GW}\simeq (1-2)\times10^3~{\rm s} \simeq~(3-6)~\tau$. The addition of magnetic dipole losses speeds-up the process, so that the star reaches the stationary football configuration somewhat earlier (Fig. \ref{fig2}, lower panel, black-solid line). After the end of the secular evolution, we do not know what the fate of the bar is. As \citet{LaiShapiro95}
have underlined, while the star approaches a Dedekind ellipsoid the gravitational evolution
timescale increases, eventually becoming comparable to the viscous dissipation one. When this
happens, $\bar{\cal C}$ is not conserved anymore and the star is expected to be driven along a
nearly-Dedekind sequence to become a Maclaurin spheroid, since this is the only final state
that does not radiate GWs or dissipate energy viscously. The addition of magnetic dipole
losses would speed up such evolution, and further spin-down the final Maclaurin state.
We thus expect to have $\Omega_{eff}$ decreasing at some point after the constant$-\bar{\cal C}$
evolution, with the dipole luminosity $L_{dip}$ also decreasing accordingly. Correspondingly,
the energy injected into the fireball will start decreasing (eventually entering in the $q<-1$ phase,
see Sec. \ref{sec1}), and the afterglow plateau is expected to end, with the light curve
turning back to the temporal decay expected in the absence of continuous energy injection.
In view of these considerations, and for the purpose of this paper, we have limited our discussion to show
that the properties of the electromagnetic plateau, even assuming this suddenly ends after the
constant$-\bar{\cal C}$ evolution, are in agreement with those typically observed in GRBs.

\section{Conclusion}
\label{sec6}

We have discussed a possible scenario where a newly formed magnetar is left over 
after a GRB explosion, and explored the hypothesis of its being subject to a secular 
bar-mode instability, including in the spin down the contributions of both radiative 
losses by magnetic dipole emission and by GWs. Following the analytical 
treatment of \citet{LaiShapiro95}, we have shown that for reasonable values of the 
physical parameters, the typical properties of GRB afterglow X-ray plateaus may be 
reproduced. A consequence of this is that, on the relatively long timescale $10^3-10^4$ s 
of the electromagnetic plateau, the advanced LIGO/Virgo interferometers may detect a 
corresponding GW signal up to $d\sim 100$~Mpc, by carrying out matched searches. Such a 
signal would be associated with an afterglow light curve plateau from a long 
sub-luminous GRB, or from a short GRB, with isotropic energy $\lesssim 10^{50}$~erg, 
which is typical of most nearby GRBs detected. For the more energetic GRBs, a 
bar-mode GW signal may be detected without a visible plateau in the afterglow. 

In conclusion, although there are considerable uncertainties about the evolutionary 
path of newborn magnetars, our analysis indicates that the scenario proposed here 
is a plausible and interesting possibility, leading to an efficient GW emission 
process which is accompanied by a distinctive  electromagnetic signature. Thus,
in view of the impending commissioning of the advanced LIGO and Virgo, we consider
that it would be highly worthwhile to test this possibility through matched
electromagnetic-GW data searches.

\acknowledgments
We are grateful to Benjamin Owen for important comments and valuable suggestions on 
this scenario, and for helping improve the manuscript.
AC thanks Fulvio Ricci for crucial support during this project, Giovanni Montani 
for important discussions, Cristiano Palomba for very helpful suggestions, and Christian D. Ott for useful comments.
This work was supported by the ``Fondazione Angelo della Riccia'' - bando A.A. 2007-2008 
\& 2008-2009 (AC), and by NSF PHY-0757155 \& NASA NNX08AL40G (PM). AC gratefully 
acknowledges the support of the Penn State Institute for Gravitation and the Cosmos (IGC).

\end{document}